\documentclass[final]{aipproc}

\def\gs{\mathrel{\raise0.35ex\hbox{$\scriptstyle >$}\kern-0.6em
\lower0.40ex\hbox{{$\scriptstyle \sim$}}}}
\def\ls{\mathrel{\raise0.35ex\hbox{$\scriptstyle <$}\kern-0.6em
\lower0.40ex\hbox{{$\scriptstyle \sim$}}}}

\layoutstyle{6x9}

\SetInternalRegister\hbadness{8000} 

\newcommand\doingARLO[2][]{
  \ifx\mmref\undefined #1\else #2\fi
}

\begin{document}

\title 
      [A robust sample of submillimetre galaxies]
      {A robust sample of submillimetre galaxies: constraints on the prevalence
       of dusty, high-redshift starbursts}


\author{R.\ J.\ Ivison}{
  address={Astronomy Technology Centre, Royal Observatory, Blackford Hill,
             Edinburgh EH9 3HJ},
  altaddress={Institute for Astronomy, University of Edinburgh, Blackford Hill,
             Edinburgh EH9 3HJ}
}

\copyrightyear  {2005}

\begin{abstract}
We develop and apply a dual-survey extraction technique to published submm
images of the Lockman Hole using the SCUBA and MAMBO bolometer arrays. Cut
above 5$\sigma$, this catalogue of submm galaxies (SMGs) is significantly more
robust than previous samples, typically selected above 3.0--3.5$\sigma$, with a
much-reduced likelihood of real, faint SMGs (beneath and around the confusion
limit) entering the catalogue via superposition with noise. Our selection
technique yields 19 SMGs of which we expect at most two to be due to chance
superposition of SCUBA and MAMBO noise peaks. The flux limit of the sample
($\sim$5\,mJy at 0.85\,mm), which is sensitive to luminous, dusty galaxies at
extreme redshifts, is optimally matched to a deep 1.4-GHz image (4.6\,$\mu$Jy
beam$^{-1}$ rms) which probes starbursts in the $z\ls\rm 4$ regime. A high
fraction of these robust SMGs ($\sim$80\%) have radio counterparts which, given
the probable $\sim$10\% contamination by spurious sources, suggests that very
distant SMGs ($z\,\gs$\,3--4) are unlikely to make up more than $\sim$10\% of
the bright SMG population. Finally, the accurate radio positions are used to
test the accuracy of positions determined in the submm, with important
implications for upcoming surveys with SCUBA-2.
\end{abstract}

\date{\today}

\maketitle

\section{Introduction}


Submm surveys pick up dusty objects at extreme redshifts, galaxies that drop
out of surveys at shorter and longer wavelengths due to obscuration and
unfavourable $K$ corrections, respectively. The first cosmological surveys
using SCUBA and MAMBO quickly and radically changed the accepted picture of
galaxy formation and evolution, moving away from the optocentric view of the
last century. The discovery of so-called `SCUBA galaxies' \cite{Sm97} was
greeted with surprise due to the remarkable evolution in the dusty, starburst
galaxy population implied by such a large source density at the flux levels
accessible to the first generation of bolometer arrays \cite{B99}. Excitement
was replaced by pessimism with the first efforts to study SMGs at optical/IR
wavelengths: early reports, backed by a study in the HDF-N \cite{H98},
suggested that the majority of the submm population had no plausible optical
counterparts. Attention was diverted to various redshift engines and broadband
photometric techniques \cite{A03,W04}. As a result, only a handful of detailed
studies were attempted, often for extreme, non-representative galaxies.

Recent progress has largely been the result of radio imaging of submm survey
fields, though early radio follow-up data were not well matched to the depth of
the submm imaging and only modest detection rates were achieved. Once
sufficient depths were attained \cite{I02}, radio maps were able to pinpoint
half of the brightest known SMGs to $\sim$0.3$''$ and, combined with the submm
flux density, provide a rough estimate of redshift. Radio data also enabled
some refinement of submm samples, increasing the detection fraction to two
thirds of SMGs at 0.85-mm flux levels in excess of $\sim$5\,mJy. With positions
in hand, these bright SMGs were found to be a diverse population --- some
quasar-like, with broad lines and X-ray detections \cite{I98,K03}, some
morphologically complex \cite{I00}, some extremely red
\cite{Sm99,I02,W03b,D04}, some with the signatures of obscured AGN and/or
superwinds \cite{Sm03}.

Spectroscopic redshifts have been extremely difficult to determine, requiring a
significant investment of time on 10-m telescopes with highly efficient
spectrometers. The first systematic survey was undertaken by Chapman et al.\
\cite{C03,Sw04,C05}. The median redshift is now known to be $\sim$2.2 for
$S_{\rm 0.85mm} \ge \rm 5$-mJy galaxies selected using SCUBA and pinpointed at
1.4\,GHz. A similar survey with a modern red-sensitive spectrometer has yet to
be attempted. We must therefore accept that present knowledge may be biased by
wavelength coverage and by the need to identify targets via radio imaging
(which does not benefit from the same $K$ correction as submm observations and
is thought to impose an artificial redshift cut-off at $z\,\sim$\,3--4).

Accurate redshifts have facilitated the first systematic measurements of
molecular gas mass for SMGs ($\sim$10$^{11}$\,M$_{\odot}$) via observations of
CO \cite{N03,G05}, as well as constraints on gas reservoir size and dynamical
mass \cite{T05}. The results agree with earlier work on what were thought to be
less representative systems \cite{F98,F99}, suggesting that SMGs are massive
systems and providing some of the strongest tests of galaxy-formation models to
date \cite{G03}.

Despite this progress, a detailed understanding of SMGs remains a distant goal.
Confusion currently limits our investigations to the brightest quartile of the
submm population (though surveys through lensing clusters have provided a
handful of sources more typical of the faint population that dominates the
cosmic background \cite{Sm02}). We must also recall that selection biases have
potentially skewed our understanding --- limited coverage of red and IR
wavelengths in spectroscopic surveys, and a third of SMGs remain undetected in
the radio and thus untargeted by existing spectroscopic campaigns.

Here, we present a robust sample of bright SMGs selected using SCUBA and MAMBO
in one of the `8-mJy Survey' regions: the Lockman Hole. Our goal is to provide
a bright sample which we would expect to detect in well-matched radio imaging
($\sigma_{\rm 0.85mm}/\sigma_{\rm 1.4GHz} \sim\rm 500$) whilst minimising, so
far as is practicable, the possibility that sources are spurious or anamalously
bright. We may thus determine the true fraction of radio drop-outs ($z>\rm 4$)
amongst SMGs, as well as practical information such as the intrinsic positional
uncertainty for SMGs in the absence of radio/IR counterparts.

\section{Sample selection}

Our objective here is to avoid one of the most common criticisms levelled at
submm surveys --- the potentially large number of spurious sources present in
submm catalogues. This has limited our ability to address the true recovery
fraction in the radio, and hence the corrections that must be made to the
redshift distributions that are used to determine star-formation histories and
galaxy-formation models.

Greve et al.\ \cite{G04} presented a 1.2-mm survey of the ELAIS N2 and Lockman
Hole regions, centred on the coordinates mapped at 0.85\,mm by SCUBA in the
`8-mJy Survey' \cite{S02}. Mortier et al.\ \cite{M05} present a refined
analysis of SCUBA data for a subset of the Lockman Hole MAMBO image.

Greve et al.\ argue that several maps with low signal-to-noise ratio (SNR) of
the same region, with only marginal differences in frequency, produce several
visualisations of essentially the same sky, tracing the same population of
luminous, dusty galaxies. We adopt this philosophy here, starting from the
1.2-mm catalogue of Greve et al., looking for matches at 0.85\,mm from Mortier
et al., substituting SMGs from Scott et al.\ where blends are evident in the
Mortier et al.\ catalogue.

Existing surveys have typically employed a SNR threshold of 3.0--3.5 and false
detections are dominated by `flux boosting', possibly at the 10--40\% level
\cite{S02}. To realise a significant improvement, reducing the probability of
an SMG being spurious, with a lower chance of flux boosting, we must limit our
sample to higher SNRs. Initially, we take those sources that have a combined
statistical significance equivalent to a signal-to-noise ratio (SNR) in excess
of 4.5 in a map with Gaussian noise properties, leaving aside the potential
clustering of SMGs \cite{B04}. We restrict ourselves to the Lockman Hole
because these data are of better quality. We then refine the selection criteria
by comparing the properties of this sample with catalogues of spurious sources
generated by the introduction of arbitrary positional offsets.

%
%
\setcounter{table}{0}
\begin{table}
\begin{tabular}{lcccccccccc}
Name&\multicolumn{2}{c}{Position at 0.85mm}&$S_{\rm 0.85mm}$&S/N&\multicolumn{2}{c}{Position at 1.2mm}&$S_{\rm 1.2mm}$&S/N&Sep&Final\\
        &$\alpha_{\rm J2000}$&$\delta_{\rm J2000}$&/mJy&&$\alpha_{\rm J2000}$&$\delta_{\rm J2000}$&/mJy&&&S/N\\
        &h m s&$^{\circ}\ '\ ''$&&&h m s&$^{\circ}\ '\ ''$&&&$''$&\\
&&&&&&&&&\\
LH-1200.001 = LE850.02      &10:52:38.6&+57:24:38&10.9\,$\pm$\,2.1&5.1&10:52:38.3&+57:24:37&4.8\,$\pm$\,0.6&8.0&2.1&$>$8\\
LH-1200.002                 &10:52:38.6&+57:23:19&5.2\,$\pm$\,2.0 &2.6&10:52:38.8&+57:23:21&4.1\,$\pm$\,0.6&6.8&3.6&7.6 \\
LH-1200.003 = LE850.14      &10:52:04.2&+57:27:01&10.5\,$\pm$\,2.0&5.2&10:52:04.1&+57:26:57&3.6\,$\pm$\,0.6&6.0&3.4&$>$8\\
                                        &          &         &                &   &10:52:03.9&+57:27:10&1.3\,$\pm$\,0.6&2.2&   &    \\
LH-1200.004                 &          &         &                &   &10:52:57.0&+57:21:07&5.7\,$\pm$\,1.0&5.7&   &5.7 \\
LH-1200.005 = LE850.01      &10:52:01.5&+57:24:43&8.6\,$\pm$\,1.1 &8.0&10:52:01.3&+57:24:48&3.4\,$\pm$\,0.6&5.7&5.1&$>$8\\
LH-1200.006 = LE850.16      &10:52:27.6&+57:25:17&7.0\,$\pm$\,1.6 &4.3&10:52:27.4&+57:25:15&2.8\,$\pm$\,0.5&5.6&2.6&7.4 \\
LH-1200.007                 &10:52:03.4&+57:18:12&5.8\,$\pm$\,1.3 &4.3&10:52:04.7&+57:18:12&3.2\,$\pm$\,0.7&4.6&10.4&6.4\\
LH-1200.008                             &10:51:40.7&+57:19:53&4.6\,$\pm$\,1.6 &2.8&10:51:41.9&+57:19:51&4.1\,$\pm$\,0.9&4.6&9.8&5.5\\
LH-1200.009                &10:52:28.2&+57:22:17&6.3\,$\pm$\,1.8 &3.5&10:52:27.5&+57:22:20&3.1\,$\pm$\,0.7&4.4&6.3&6.0\\
LH-1200.010 = LE850.06     &10:52:30.4&+57:22:13&10.8\,$\pm$\,2.4&4.6&10:52:29.9&+57:22:05&2.9\,$\pm$\,0.7&4.1&9.5&6.4\\
LH-1200.011 = LE850.03      &10:51:58.3&+57:18:00&5.0\,$\pm$\,1.3 &3.9&10:51:58.2&+57:17:53&2.9\,$\pm$\,0.7&4.1&7.6&6.0\\
                                        &          &         &                &   &10:51:57.6&+57:18:05&1.6\,$\pm$\,0.7&2.3&   &   \\
LH-1200.012 = LE850.18      &10:51:55.9&+57:23:13&4.3\,$\pm$\,1.0 &4.2&10:51:55.4&+57:23:10&3.3\,$\pm$\,0.8&4.1&5.2&6.2\\
LH-1200.014 = LE850.08      &10:51:59.6&+57:24:21&4.8\,$\pm$\,1.1 &4.3&10:52:00.0&+57:24:24&2.4\,$\pm$\,0.6&4.0&4.9&6.2\\
LH-1200.017                 &10:51:22.5&+57:18:42&14.6\,$\pm$\,4.1&3.5&10:51:21.4&+57:18:40&4.8\,$\pm$\,1.3&3.7&8.8&5.3\\
LH-1200.019                             &10:51:27.8&+57:19:47&6.7\,$\pm$\,2.3 &2.9&10:51:28.3&+57:19:46&4.0\,$\pm$\,1.1&3.6&4.5&5.0\\
LH-1200.022                             &10:52:03.3&+57:15:37&7.3\,$\pm$\,2.2 &3.3&10:52:03.0&+57:15:46&2.8\,$\pm$\,0.8&3.5&9.1&5.0\\
LH-1200.042 = LE850.29      &10:52:16.2&+57:25:05&7.2\,$\pm$\,1.5 &4.7&10:52:16.0&+57:25:06&1.6\,$\pm$\,0.5&3.2&1.9&6.0\\
LH-1200.096 = LE850.07     &10:51:51.4&+57:26:38&6.7\,$\pm$\,1.7 &4.0&10:51:51.4&+57:26:40&1.6\,$\pm$\,0.6&2.7&4.8&5.1\\
LH-1200.104 = LE850.27     &10:51:53.9&+57:18:38&6.3\,$\pm$\,1.2 &5.2&10:51:53.7&+57:18:39&2.1\,$\pm$\,0.8&2.6&2.1&6.2\\
\end{tabular}
\caption{Combined sample of MAMBO/SCUBA sources in the Lockman Hole.}
\end{table}

%
%
\setcounter{figure}{0}
\begin{figure}
\includegraphics[height=.35\textheight]{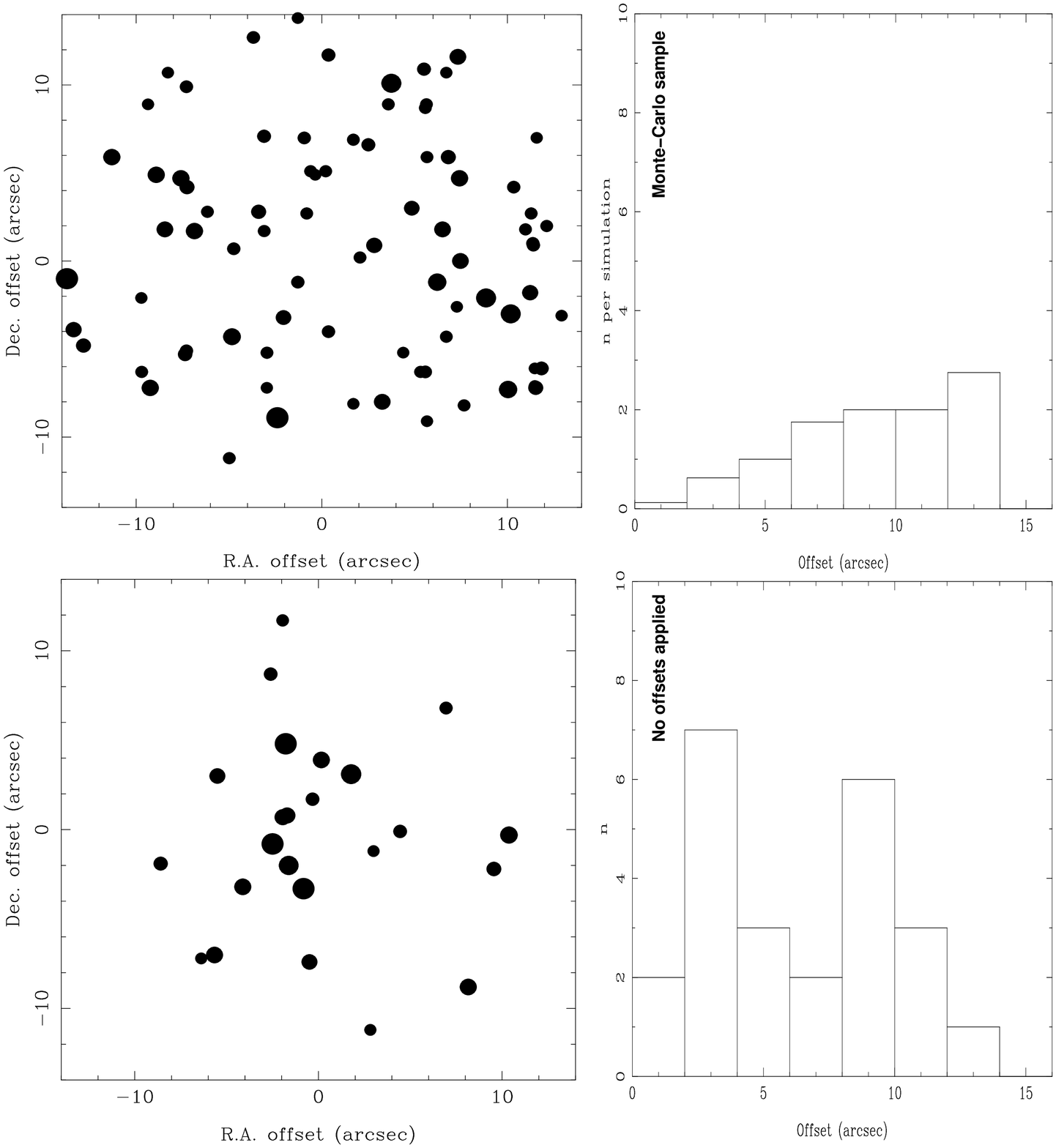}
\includegraphics[height=.35\textheight]{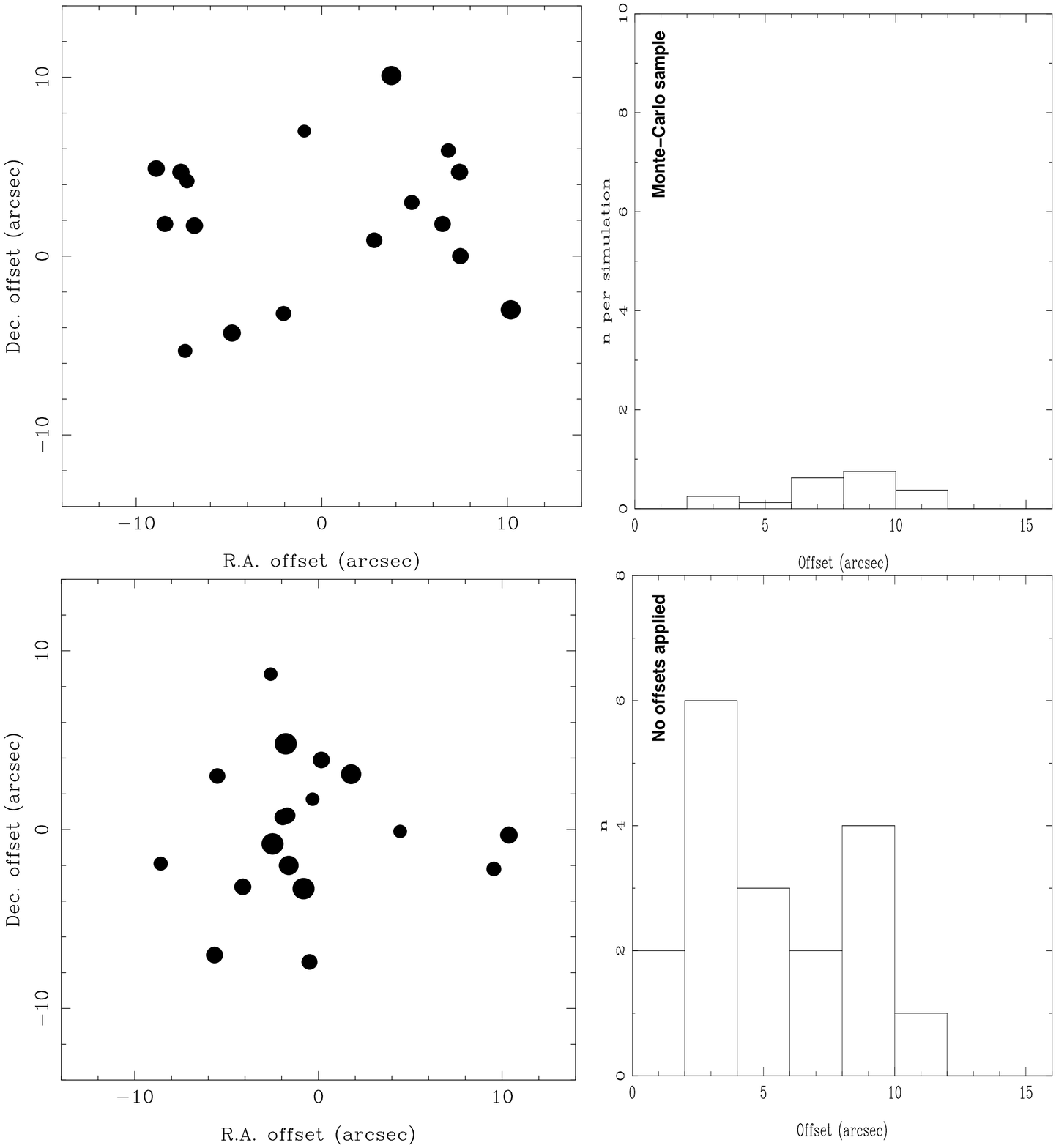}
\caption{{\em Left:} Positional offsets between SCUBA and MAMBO peaks for
spurious ({\em top}) and real ({\em bottom}) catalogues with a combined
significance above 4.5$\sigma$. Larger symbols represent a higher combined
SNR. Spurious sources were generated by offsetting MAMBO catalogue
positions. Histograms show the radial offset between SCUBA and MAMBO peaks for
spurious ({\em top}, normalised to one simulation) and real ({\em bottom})
catalogues. {\em Right:} Positional offsets between SCUBA and MAMBO peaks for
spurious ({\em top}) and real ({\em bottom}) catalogues, after application of
selection filters described in the text.}
\end{figure}

%
%
\setcounter{figure}{1}
\begin{figure}
\includegraphics[height=.4\textheight]{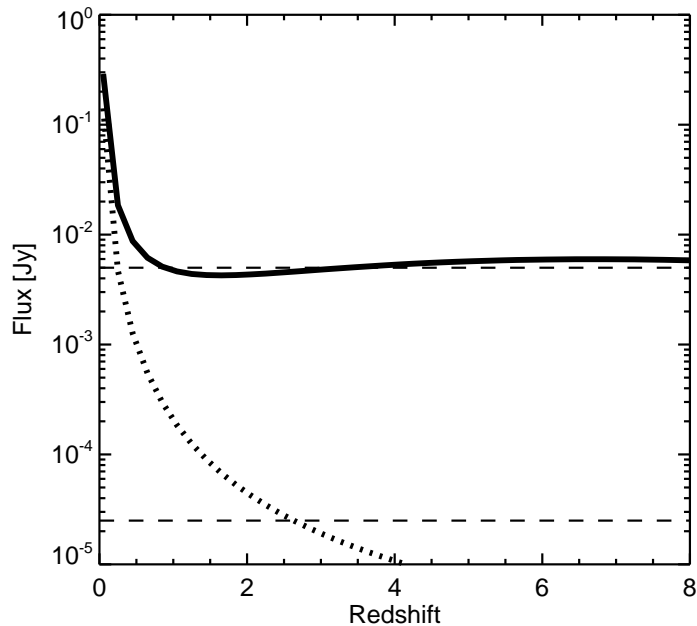}
\caption{Flux density versus redshift for an Arp\,220 spectral energy
distribution, at 5$\times$ Arp\,220's luminosity. The variation with
redshift is shown for flux densities at 1\,mm and 1.4\,GHz (solid and
dotted lines, respectively), with the 5-$\sigma$ detection limits indicated
as dashed horizontal lines. As is widely known, mm and submm surveys are
sensitive to dusty starbursts out to extreme redshifts ($z\gs\rm 6$) but
the deepest current radio surveys lose sensitivity to these galaxies at
$z\,>\,$3--4.}
\end{figure}

To accomplish this dual-survey extraction, where there are numerous
permutations of SNRs that yield the desired SNR, we have developed a simple
algorithm. We take the 2-$\sigma$ catalogue from the MAMBO image, extracted as
described by Greve et al.\ \cite{G04}. We then search for SCUBA sources within
14$''$ (roughly the area of five beams), applying a correction for separation
and checking for a combined significance above my initial threshold,
4.5$\sigma$. We include sources that exceed 4.5$\sigma$ in either dataset, as
long as there is a valid reason why the source is not seen in the other image,
and sources with a combined probability exceeding 4.5$\sigma$ with a minimum
2-$\sigma$ detection in both images, within 14$''$ of one another. To account
for the distance between sources, we divide the probabilities by the number of
beams inside a circle with a radius equal to the separation of the sources.

The noise properties of submm images are not Gaussian due to the miriad of
real, faint sources near the confusion limit. The most important effect of
increasing the SNR threshold should be to drastically reduce the false
detections due to faint SMGs that have been boosted above the detection
threshold by noise --- flux boosting \cite{S02,G04}. However, with so many
$\ge$2-$\sigma$ sources in the catalogues, it is important to investigate how
many SMGs may contaminate the combined sample due to random coincidence of
faint MAMBO and SCUBA peaks. To this end, we performed simulations, offsetting
the MAMBO sample by $\pm$30$''$ and $\pm$45$''$ (in R.A.\ and Dec.). Each
simulation typically yields ten SMGs --- far too many to hope that a
statistically robust catalogue would emerge from the process. The results of
these simulations, and for the real samples, are illustrated in the left-hand
panels of Fig.~1. The left-hand panels show the scatter of R.A.\ and Dec.\
offsets between each MAMBO and SCUBA pair, with symbol size coded to match the
SNR of the highest peak. The right-hand panels show histograms of $\sqrt
(\Delta \alpha^2 + \Delta \delta^2)$ --- the radial offset, $r$. The simulated
catalogues clearly yield a very different distribution of offsets from the
position-matched data, exactly the $n \propto r^2$ form expected for randomly
scattered peaks, and without the expected concentration of high-SNR pairs at
low $r$.

Using the information in these plots we have taken a number of approaches to
minimise the number of spurious SMGs in our sample. Most of the sources in the
simulated catalogues are the result, perhaps unsurprisingly, of near-coincident
low-SNR peaks. We found that raising the minimum catalogue threshold to
$\ge$2.5$\sigma$ reduced the number of false detections whilst having no effect
on the real catalogue. Lowering the search radius to 11$''$ reduced the number
of false positives in line with the ratio of the respective search areas,
removing only one source from the real catalogue. Insisting that the higher of
the two peaks is $\ge$3.5$\sigma$ removes a further quarter of the simulated
SMGs, whilst reducing the real catalogue by half that amount.  Finally, we
increased the combined theshold to $\ge$5$\sigma$ to further reduce the number
of rogue sources. The effects of this approach on the SCUBA--MAMBO positional
offsets are shown in the right-hand panels of Fig.\ 1. We are left with a
situation where the final catalogue consists of 19 SMGs of which we expect at
most two to be the result of coincidence. It is clear that the effect of
confusion is less apparent than in previous catalogues, without the sample size
shrinking dramatically, though the sample is not entirely immune. We list the
resulting 19 sources in Table~1.

Note that all five of the SCUBA sources detected individually above 5$\sigma$
are also found in the MAMBO image with a SNR of at least 2.0, though one is
below 2.5$\sigma$ and has thus been excised from my final sample. Of the eight
$\ge$5-$\sigma$ MAMBO sources, only one was not seen by SCUBA at
$\ge$2.5$\sigma$: LE-1200.4, which lies off the SCUBA map, to the East, but was
subsequently detected by Scott et al.\ (in preparation).

\section{Identification of the SMGs in the radio}

The process of identifying counterparts to SMGs has been refined in a series of
studies \cite{I98,I00,I02,I04,Sm99,Sm00,W03a,W03b,P05}. The most effective
methods employ a combination of radio, optical and IR imaging, searching for
the red rest-frame optical light expected of a dust-enshrouded galaxy and the
synchrotron emission expected of a starburst or radio-loud active galaxy. Radio
sources and EROs are sufficiently rare that associations of either with SMGs
can be viewed as robust in most cases.

%
%
\setcounter{figure}{2}
\begin{figure}
\includegraphics[height=.65\textheight]{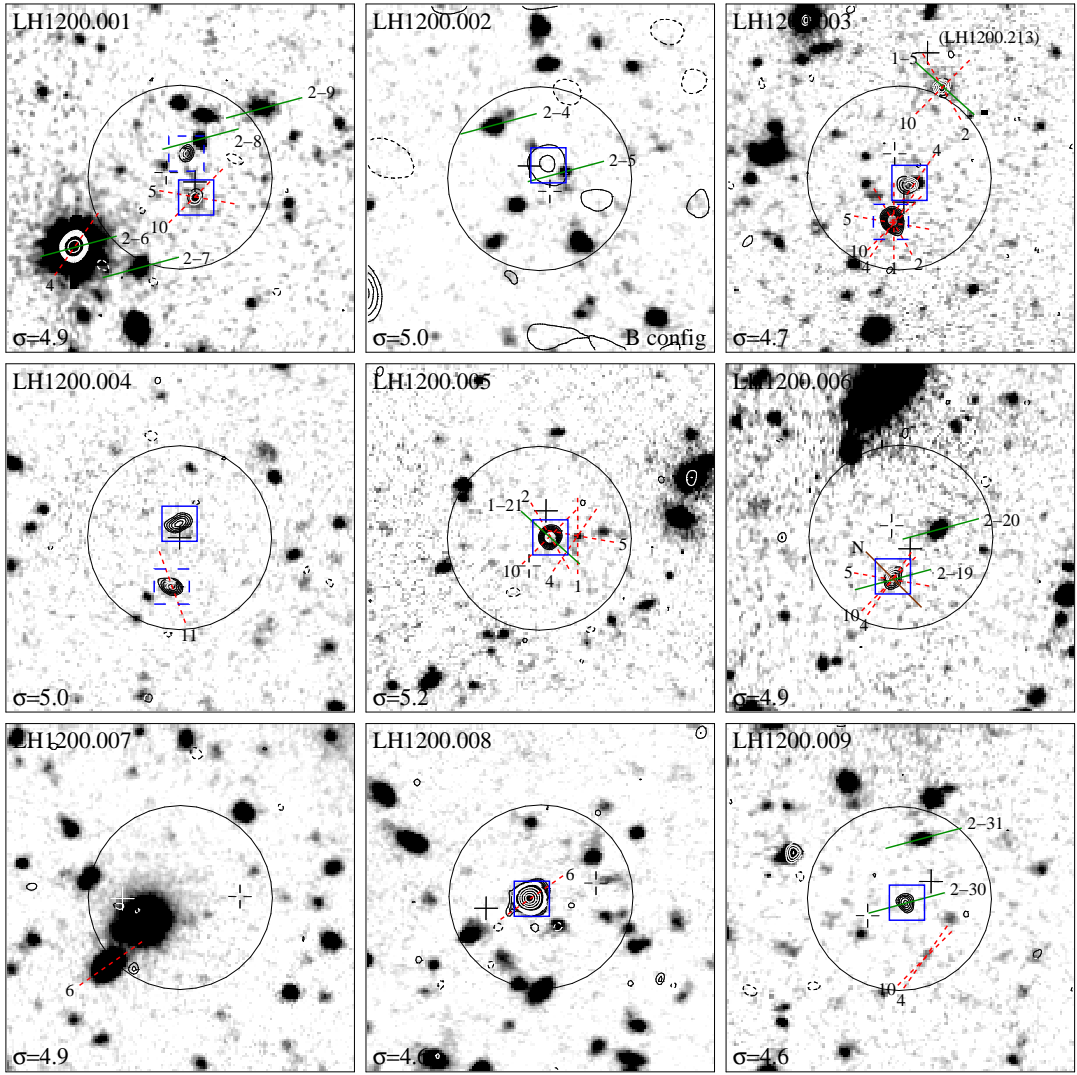}
\caption{Postage stamps ($30'' \times 30''$) of the fields surrounding the
first nine $>$5$\sigma$ SMGs in the Lockman Hole. Optical ($R$) data are shown
as a grayscale upon which 1.4-GHz contours are plotted.  Open crosses mark
SCUBA galaxies; solid crosses mark MAMBO galaxies; 8$''$-radius circles mark
the average positions; dashed lines represent the position of GMOS, LRIS and
NIRSPEC slits \cite{C03,Sw04,C05}; squares mark radio counterparts with $P<\rm
0.05$, solid squares being the most probable in each case.}
\end{figure}

Fig.~2 shows how the flux density of a luminous, dusty starburst varies with
redshift at both 1\,mm and 1.4\,GHz, adopting the SED of Carilli \& Yun
\cite{CY99}. To identify a $z\rm \le 3$ 5-mJy SMG (the average of the 0.85- and
1.2-mm flux densities) in a 1.4-GHz image, it is clear that the radio data must
achieve a sensitivity of at least 5\,$\mu$Jy beam$^{-1}$.

Even at this radio sensitivity, the lack of a robust radio identification could
have at least five origins: i) the SMG could be spurious; ii) the SMG could be
flux boosted significantly; iii) the radio/far-IR emission could be
significantly larger than the 1.4-GHz synthesised beam \cite{I02}; iv) the
characteristic dust temperature could be low \cite{C04}; or, most
interestingly, v) the SMG could lie at $z\,\gs$\,3--4 \cite{E03}.  It is
difficult, given the quality of existing samples, and the mis-matched
radio/submm datasets, to determine which of these are important, though Ivison
et al.\ \cite{I02} showed that (i) affects {\em at least} 15\% of a 3.5$\sigma$
sample.

Deep, high-resolution, wide-field radio images were obtained at the Very Large
Array and employed to pinpoint the SMGs. The data used here are described in
detail by Ivison et al.\ \cite{I02}. Fig.~3 shows postage stamps around the
positions given in Table~2 (see the printed version of this document) for the
nine most significant SMGs.  A radio source peaking at $\ge$\,4\,$\sigma$ in
the 1.4$''$ or smoothed images, with an integrated flux density in excess of
15\,$\mu$Jy, is considered a {\it robust} detection. Fainter sources, where the
definition is relaxed to only the integrated flux, were also catalogued.

For each SMG we have searched for a potential radio (1.4-GHz) counterpart out
to a radius of 8$''$ from the mid-point of the 0.85- and 1.2-mm emission. This
relatively large search area (201$''^2$ around each source) represents a
$\sim$99\% positional confidence region (see later) and should ensure that few
real associations are missed.

To quantify the formal significance of each of the potential submm/radio
associations, we have used the method of Downes et al.\ \cite{D86} to correct
the raw Poisson probability, $P$, that a radio source of the observed flux
density could lie at the observed distance from the SMG for the number of ways
that such an apparently significant association could have been uncovered by
chance.

Of the 19 SMGs in the sample, only four lack robust radio counterparts. The
flux densities and positions of all candidate radio counterparts are listed in
Table~2 (see the printed version of this document), along with the search
positions. Of the four sources which have more than one potential radio
counterpart, we find that the correct identification is never statistically
obvious. The formal probability of the second candidate association occurring
by chance is low, $P \le\rm 0.05$. The obvious interpretations of such multiple
statistical associations are either gravitational lensing (implausible in most
of the cases here), or clustering of star-forming objects/AGN at the source
redshift (already proven in several cases --- \cite{L02,C05}); another
possibility is that sources with multiple radio counterparts are boosted into
bright submm catalogues by virtue of comprising multiple, faint, physically
unrelated SMGs, i.e.\ by conventional confusion.

This calculation has yielded statistically robust radio counterparts for 15 of
the 19 SMGs. The plausibility of this figure can be checked by noting the ratio
of areas inside and outside the circles in Fig.~3, 4.5:1. Ten random `field'
radio sources are detected in the outer areas so, given that $\mu$Jy radio
sources are expected to be over-dense around SMGs, we expect only a handful (at
most 2--3) of the counterparts to be spurious. Based on the $P$ values given in
Table~2 (see the printed version of this document), the most likely candidates
for spurious associations are LH-1200.007 and LH-1200.019, leaving 15/19 robust
radio identifications.

%
%
\setcounter{figure}{3}
\begin{figure}
\includegraphics[height=.4\textheight]{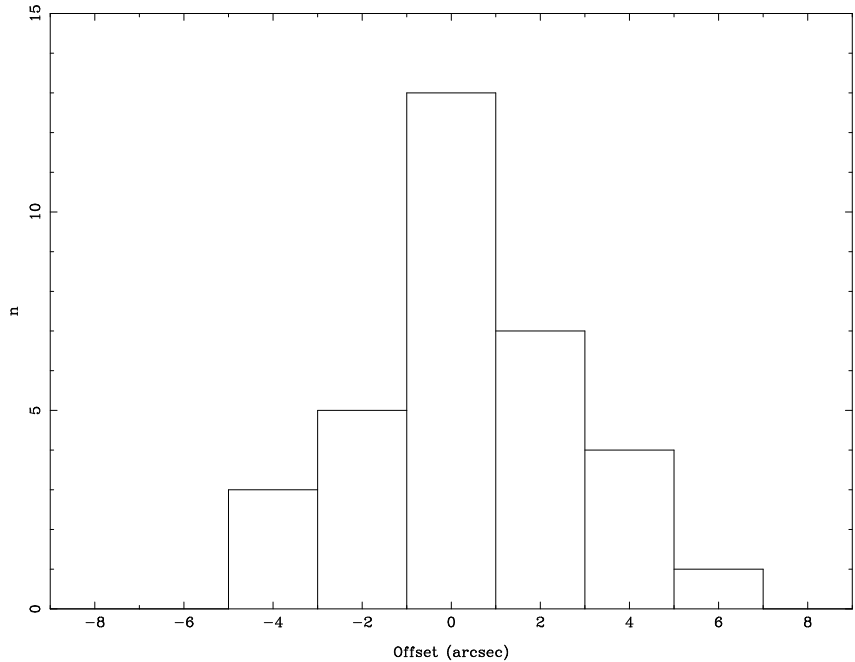}
\caption{Histogram of positional offsets, in R.A.\ and Dec., between the
(sub)mm centroid and the position of the most likely radio counterpart. A
Gaussian fit to the distribution yields a {\sc fwhm} of 5.2 $\pm$ 1.2$''$,
which translates into a 1-$\sigma$ uncertainty of 2.2$''$ for our (sub)mm
positions.}
\end{figure}

\subsection{Implications for submm positional uncertainty}

Fig.~4 shows a histogram of R.A.\ and Dec.\ positional offsets between the
mm/submm centroids and the most likely radio counterparts. A Gaussian fit to
the distribution yields a {\sc fwhm} of 5.2 $\pm$ 1.2$''$, which translates
into an r.m.s.\ separation of 5.2$''$/2.354 = 2.2$''$ between the (sub)mm and
radio positions. The sample can thus be employed to re-calibrate the
rule-of-thumb relationship between positional accuracy, beam size and SNR
(typically, SNR\,$\sim$\,6 here). We must acknowledge a mild circularity to the
logic, given that positional offsets have been used to calculate $P$, though
this may be offset by the lack of correction for radio sources in
the field.

The conventional positional uncertainty, $\sigma_r$, occurs where the
distribution of radial offsets peaks; this is the same as the the r.m.s.\
separation deduced earlier. Within this radius we expect to find 39.3\% of the
population, with 86.5 and 98.9\% within 2\,$\sigma_r$ and 3\,$\sigma_r$ (from 1
-- $e^{-r^2/2\sigma_r^2}$). In the absence of radio counterparts, it would seem
from our analysis that around 39\% of SMGs can be located within a radial
distance of $\sim\theta/\rm SNR$, where $\theta$ is the {\sc fwhm} beam size,
in arcsec). Thus the 1-$\sigma_r$ positional uncertainty for 3--3.5$\sigma$
SCUBA-selected SMGs is 4--5$''$, cf.\ the rule of thumb quoted by Hughes et
al.\ \cite{H98}. The most secure SMGs, at $\sim$10$\sigma$, representative of
those expected in upcoming, confusion-limited, wide-field (tens of square
degrees) 0.85-mm surveys using SCUBA-2, with lower significance, simultaneous
detections at 0.45\,mm, will be located with a uncertainty of $\sigma_r \rm
\sim 1''$, at which level the precision of the telescope pointing and SCUBA-2
flatfield may become important. Assuming these sources of uncertainty can be
minimised, the current requirement for deep radio coverage to identify
counterparts and enable spectroscopic follow-up may not be as urgent,
particularly if $\rm 3''\times 3''$ deployable integral field units are
employed (e.g.\ KMOS \cite{S03}).

\begin{theacknowledgments}
Many thanks to my trusty collaborators: Thomas Greve, James Dunlop, Mark
Swinbank, Jason Stevens, Angela Mortier, Steve Serjeant, Frank Bertoldi, Ian
Smail, Andrew Blain and Scott Chapman.
\end{theacknowledgments}

\doingARLO[\bibliographystyle{aipproc}]
          {\ifthenelse{\equal{\AIPcitestyleselect}{num}}
             {\bibliographystyle{arlonum}}
             {\bibliographystyle{arlobib}}
          }

\end{document}